# Опис інтерференції в адронних і партонних зливах методом Лапласа

Потієнко О. С., Шарф І. В., Чудак Н. О., Дьомін Д. С., Шелковенко І. О., Меркотан К. К.

*Державний університет «Одеська політехніка»*
*Проспект Шевченка 1, Одеса, 65044, Україна*

**Анотація**

В роботі показано, що застосування методу Лапласа для деревних діаграм дозволяє розрахувати інтерференційні доданки, які виникають в наслідок підсумування по перестановках частинок в кінцевому стані розсіяння. Ці доданки суттєво впливають на значення експериментально спостережуваних величин. Доведено, що деревна діаграма з однією вхідною лінією і довільною кількістю вихідних ліній має максимальне значення при рівних чотириімпульсах вторинних частинок. Це дозволяє стверджувати, що деревні діаграми непружного розсіяння описують процеси утворення партонних і адронних злив. В роботі показано необхідність врахування внесків в парціальні перерізи, які виникають за рахунок інтерференції між різними зливами.

## 1. Вступ

В нашій попередній роботі [1] ми застосували метод Лапласа [2] для аналізу інтерференційних ефектів в процесах непружного розсіяння при високих енергіях, коли утворюється велика кількість вторинних частинок. При цьому в роботі [1] розглядався лише один тип безпетльових діаграм непружного розсіяння – діаграми типу «гребінка». Проте, запропонований в роботі [1] метод врахування інтерференційних внесків може бути застосований до ширшого типу діаграм, зокрема схематично показаних на рис.1. Кожне коло позначає суму всіх можливих безпетльових (деревних) діаграм із однією вхідною лінією і заданою кількістю вихідних ліній. Приклад однієї з можливих діаграм наведено на рис. 2. Кожну таку суму будемо далі називати деревним блоком. За допомогою методу Лапласа далі буде показано, що подібні діаграми описують процес утворення декількох злив. Метою цієї роботи є вивчення інтерференційних ефектів всередині злив і між зливами.

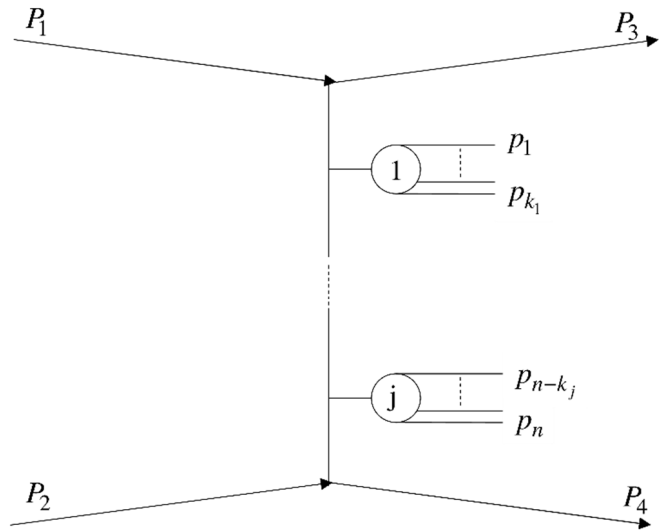

Рис.1. Приклад діаграми моделі «фі-три», що описує утворення декількох адронних струменів.

Спочатку ми проведемо розгляд на прикладі найпростішої моделі «фі-три», а в кінці роботи обговоримо застосування отриманих результатів для моделювання партонних злив в КХД.

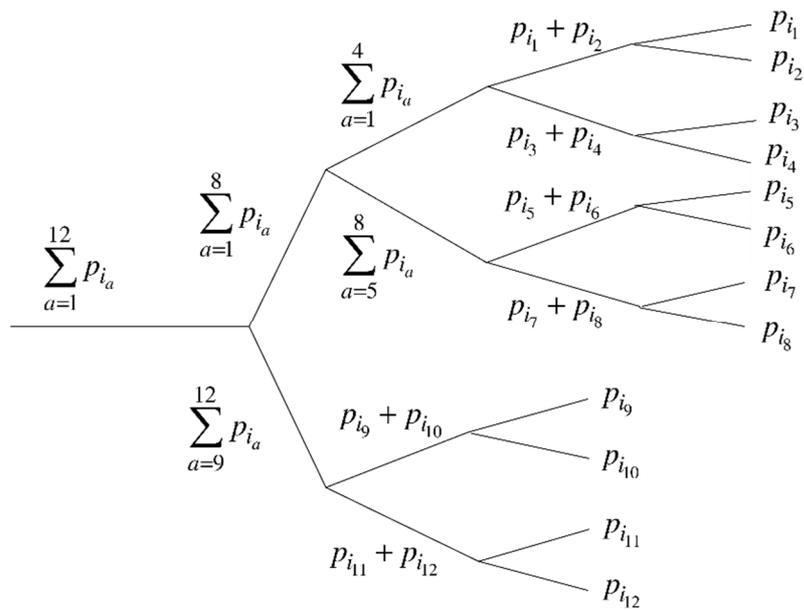

Рис. 2. Приклад типової деревної діаграми.

Зазвичай зливові процеси описують в межах моделі ДГЛАП [3-5]. Ця модель заснована на припущенні що основний внесок в переріз процесу з утворенням злив дає область, в якій поперечні до вісі зливи компоненти імпульсів сильно впорядковані [4,6]. Це мотивується тим, що при інтегруванні по цій області константа зв'язку підсилюється максимальним ступенем великого логарифму. Але ці міркування не враховують інтерференцію всередині зливи. Припущення про сильну впорядкованість поперечних імпульсів практично «виключає» цю інтерференцію з розгляду. В той же час, далі в цій роботі ми покажемо, що максимальний внесок в переріз вносить область, в якій поперечні до вісі зливи компоненти імпульсів не є сильно впорядкованими, а всі дорівнюють нулю. Врахування інтерференції всередині зливи призводить до того, що в цій області константа зв'язку підсилюється факторіально, що може бути сильніше ніж підсилення ступенями великого логарифма. Наш розгляд інтерференцій всередині злив суттєво відрізняється від наведеного в роботах [7-9], де інтерференційні ефекти зводяться лише до взаємознищення внесків областей, в яких невпорядковані кути розльоту партонів.

Застосування метода Лапласа дозволяє врахувати не тільки інтерференцію всередині злив, а й внески від інтерференційних доданків між різними зливами. Нажаль, нам не вдалося знайти роботи в яких розглядалося б питання врахування інтерференції між зливами. Далі в цій роботі ми покажемо, що врахування цієї інтерференції є важливим.

## 2. Алгоритм побудови одного деревного блока

Побудуємо за допомогою діаграми, показаної на рис.3а, всі можливі діаграми з однією вхідною і трьома вихідними лініями. Додаткова вихідна лінія на довільній діаграмі може з'явитися лише разом із вершиною, до якої вона приєднана. Оскільки мова йде про діаграми моделі «фі-три», до цієї вершини приєднуються ще дві лінії рис.3б. Якщо ми обидві ці лінії спаримо із двома зовнішніми лініями діаграми рис.3а , то отримаємо петлю. Тому маючи на меті побудувати безпетльову діаграму ми спаримо одну із цих ліній з однією із трьох зовнішніх ліній діаграми рис. 3а. Тоді за рахунок спарювання матимемо мінус одну зовнішню лінію, але одночасно додаються дві зовнішні лінії від приєднаної вершини. Таким чином, отримаємо діаграму із кількістю вихідних ліній, більшою на одиницю від кількості таких ліній у діаграмі на рис.3а. Якщо, згідно з [10, 11], спарювати лінію від вершини, що додається послідовно із кожною зовнішньою лінією діаграми рис. 3а, отримаємо всі можливі діаграми із трьома вихідними лініями. Цей процес показано на рис. 4.

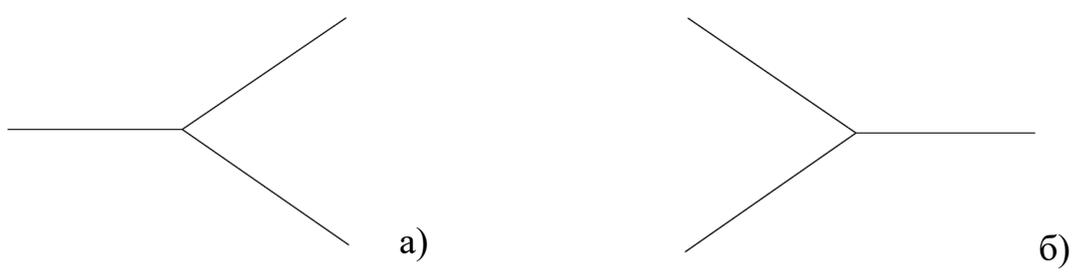

Рис. 3. Найпростіший деревний блок теорії «фі-три» (а) і приєднувана до нього вершина із трьома лініями (б).

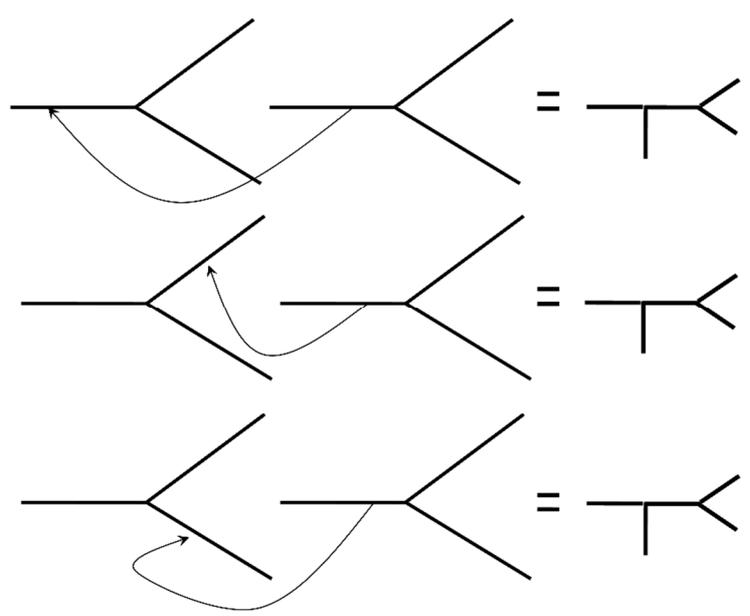

Рис. 4. Варіанти приєднання нової вершини до найпростішої діаграми.

Як видно з рис. 4, в усіх трьох випадках отримано одну й ту саму діаграму. Оскільки теорема Віка вимагає враховувати всі можливі варіанти спарювань, ми маємо врахувати внески всіх трьох діаграм або прямим підсумуванням, або введенням вагового множника. Суттєвим є той факт, що додавання однієї вихідної лінії до діаграми супроводжується додаванням однієї вершини та однієї внутрішньої лінії. Так відбувається тому, що із зовнішніх ліній вихідної діаграми, з якою спарюється лінія з доданої вершини, стає внутрішньою лінією. Описаний процес додавання вершин можна продовжувати, починаючи тепер з діаграми рис. 4.

Припустимо, що таким чином ми побудували всі деревні діаграми із $n$ зовнішніми лініями і хочемо використати ці діаграми до побудови множини всіх діаграм із $n+1$ зовнішньою лінією. Знову додавання цієї лінії можливе лише разом із вершиною, до якої приєднані ще дві лінії. Доведемо, що якщо обидві ці лінії спарити з двома зовнішніми лініями довільної діаграми попереднього кроку із $n$ зовнішніми лініями, то ми обов'язково отримаємо петлю. Дійсно, приспустимо, що нам потрібно прокласти шлях уздовж ліній діаграми від деякої вихідної лінії до вхідної. Приклад такого шляху показано стрілками на рис.5. При цьому домовимось, що дозволеними є лише такі шляхи, коли жодна лінія діаграми не проходиться двічі.

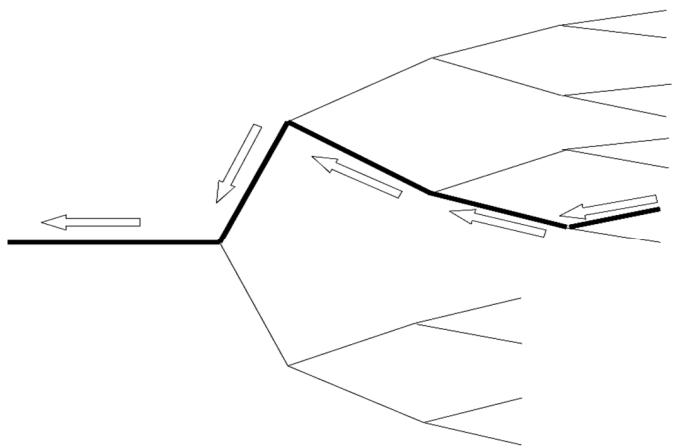

Рис.5. Приклад дозволеного шляху уздовж діаграми.

Приклад забороненого шляху, який містить лінію, що проходиться двічі, показано на рис. 6.

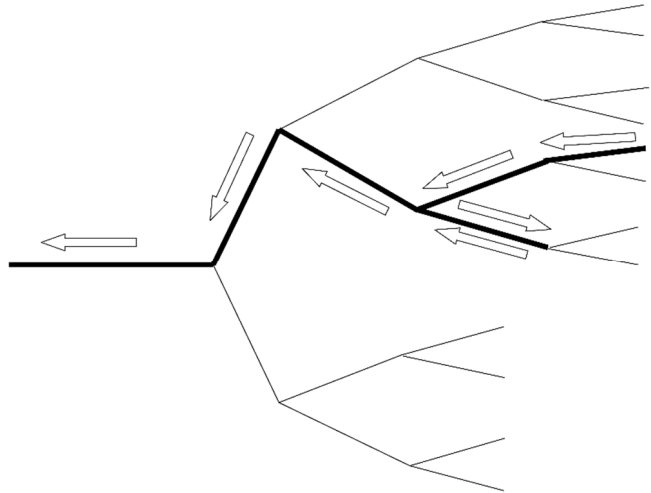

Рис. 6. Приклад забороненого шляху уздовж діаграми.

Тоді те, що діаграма не містить петель, означає, що для кожної вихідної лінії існує лише один дозволений шлях, яким можна, входячи в діаграму через вихідну лінію, вийти з діаграми через вхідну лінію. Припустимо, що додаючи до діаграми з $n$ зовнішніми лініями нову вершину, ми дві приєднані до неї лінії спаримо із двома вихідними лініями. Від кожної із цих зовнішніх ліній існував один шлях до вхідної лінії. Тому, рухаючись від доданої лінії через кожну із них, ми отримаємо два різні шляхи виходу з діаграми через вхідну лінію. Це, вочевидь, призведе до формування петлі. Тому, додаючи нову вершину до діаграми, для того, щоб не утворювалося петель, ми маємо лише спарити її з однією із зовнішніх ліній діаграми попереднього кроку, обираючи цю лінію всіма можливими способами. При цьому ця зовнішня лінія перетвориться на внутрішню і ми матимемо плюс одну внутрішню лінію.

В той же час, матимемо мінус одну зовнішню лінію, але одночасно - плюс дві зовнішні лінії від доданої вершини, тобто перехід від діаграми з $n$ зовнішніми лініями до діаграми з $n+1$ зовнішніми лініями супроводжуватиметься додаванням однієї вершини, однієї внутрішньої лінії та однієї зовнішньої лінії.

Звідси випливає важливий висновок: всі деревні діаграми з певною кількістю зовнішніх ліній мають однакову кількість вершин та однакову кількість внутрішніх ліній. Це означає, що якщо зібрати постійні множники, які відповідають вершинам і внутрішнім лініям, то отримаємо один й той самий коефіцієнт при різних діаграмах із певною кількістю зовнішніх ліній. Розглядаючи суму всіх таких діаграм, ми можемо цей спільний коефіцієнт винести за дужки.

Деревний блок, що відповідає діаграмі рис. 4.3, містить три зовнішні лінії, одну вершину і жодної внутрішньої лінії. Блок наступного кроку рис. 4.4 містить дві вершини, чотири зовнішні лінії та одну внутрішню. Оскільки на кожному кроці розглянутої вище процедури додається по одній вершині, внутрішній і зовнішній лінії, то співвідношення між кількостями зовнішніх, внутрішніх ліній і вершин зберігається. А саме, кількість внутрішніх ліній на три менша ніж зовнішніх, а кількість вершин на дві менша ніж кількість зовнішніх ліній. Відповідно, кількість внутрішніх ліній на одну менша ніж кількість вершин. Отже, для діаграми з $n$ вихідними лініями (тобто з $n+1$ зовнішніми лініями, враховуючи одну вхідну) матимемо $n-1$ вершину. Відповідно матимемо множник $\left(-i(2\pi)^4 g\right)^{n-1}$, де $g$ – константа зв'язку, що відповідає кожній вершині.

Кожній з $n-2$ внутрішніх ліній, окрім фейнманівського знаменника, який розглядатиметься далі, зіставляється множник $\left(-\dfrac{i}{(2\pi)^4}\right)$. Ще один такий множник зіставляється вхідній лінії, бо вона на відміну від вихідних відповідає віртуальній частинці. Тому маємо коефіцієнт $\left(-\dfrac{i}{(2\pi)^4}\right)^{n-1}$. З урахуванням коефіцієнту від вершин, отримаємо

$$\left(-i(2\pi)^4 g\right)^{n-1}\left(-\frac{i}{(2\pi)^4}\right)^{n-1} = (-1)^{n-1} g^{n-1} \qquad (1)$$

Окрім того, кожній внутрішній лінії, а також вхідній лінії зіставляється множник виду:

$$F = \frac{1}{m^2 - \left(\sum_a p_{i_a}\right)^2 - i\varepsilon}, \qquad (2)$$

де $m$ – маса віртуальної частинки, $\varepsilon$ – мала добавка, яка відповідає за обхід полюсів виразу (2), $p_{i_a}$ – елементи підмножини чотириімпульсів вторинних частинок сума яких переноситься уздовж лінії що розглядається.

Для різних ліній блоку кількість доданків у сумі $\sum_a p_{i_a}$ і підмножина, яку пробігають значення індексу $a$, є різними. Оскільки в кожній внутрішній лінії блоку зливаються щонайменше два чотириімпульси зовнішніх частинок, маємо

$$\left(\sum_a p_{i_a}\right)^2 \geq 4m^2. \qquad (3)$$

Це означає, що для будь-якої внутрішньої лінії відповідний знаменник в фізичній області розглядуваного процесу не обертається в нуль і кожен з цих знаменників приймає від'ємне значення. Перша обставина дає змогу наблизити $\varepsilon$ в (2) до нуля відразу, а не після закінчення розрахунків, друга - дозволяє записати замість (2)

$$F = \frac{1}{m^2 - \left(\sum_a p_{i_a}\right)^2}. \qquad (4)$$

Збираючи знаки, що відповідають кожній внутрішній лінії і вхідній лінії, отримаємо $(-1)^{n-1}$. Цей множник компенсує аналогічний множник у виразі (1). В результаті дістаємо висновку, що кожен деревний блок із $n$ вихідними лініями входить до діаграми типу рис. 1 в виді суттєво додатного множника із коефіцієнтом $g^{n-1}$ (не враховуючи множника, що походить з вершини до діаграми, до якої приєднується блок). Далі ми доведемо, що величина цього суттєво додатного множника буде максимально можливою за умови, що всі чотири імпульси, які відповідають вихідним лініям цього блоку, однакові.

## 3. Точка максимуму множника, що відповідає деревному блоку

Дійсно, знаменник виразу (4) за однакових значень чотириімпульсів і з урахуванням умови масової поверхні прийматиме значення

$$D = \left(k^2 - 1\right)m^2, \qquad (5)$$

де $k$ – кількість доданків, що утворюють суму $\sum_a p_{i_a}$, яка входить у вираз (4).

Позначатимемо далі ці імпульси індексами від 1 до $k$ і перепишемо вираз (4) у вигляді

$$F = \frac{1}{\left(k^2 - 1\right)m^2 + \triangle D}, \qquad (6)$$

де $\triangle D = \left(\sum\limits_{a=1}^{k} p_a\right)^2 - k^2 m^2$.

Врахуємо тепер, що квадрат суми чотириімпульсів, які входять в вираз для, $\triangle D$ може бути записаний у виді

$$\left(\sum_{a=1}^{k} p_a\right)^2 = \sum_{a=1}^{k}(p_a)^2 + 2\sum_{a=1}^{k-1}\sum_{b=a+1}^{k} p_a p_b = km^2 + 2\sum_{a=1}^{k-1}\sum_{b=a+1}^{k} p_a p_b. \qquad (7)$$

Відповідно, величина $\triangle D$ (6) приймає вигляд

$$\triangle D = 2\sum_{a=1}^{k-1}\sum_{b=a+1}^{k} p_a p_b - k(k-1)m^2. \qquad (8)$$

Врахуємо, що

$$\sum_{a=1}^{k-1}\sum_{b=a+1}^{k} m^2 = k(k-1)m^2, \qquad (9)$$

тоді

$$\triangle D = 2\sum_{a=1}^{k-1}\sum_{b=a+1}^{k}\left(\sqrt{m^2+\vec{p}_a^2}\sqrt{m^2+\vec{p}_b^2} - (\vec{p}_a\cdot\vec{p}_b) - m^2\right). \qquad (10)$$

Тут враховано також умову масової поверхні і позначено через $\vec{p}_a$ і $\vec{p}_a$ сукупності просторових компонент відповідних чотириімпульсів.

Зробимо далі перетворення

$$\sqrt{m^2+\vec{p}_a^2}\sqrt{m^2+\vec{p}_b^2} - m^2 = \frac{\left(m^2+\vec{p}_a^2\right)\left(m^2+\vec{p}_b^2\right)-m^4}{\sqrt{\left(m^2+\vec{p}_a^2\right)\left(m^2+\vec{p}_b^2\right)}+m^2} = $$
$$= \frac{m^2\left(\vec{p}_a^2+\vec{p}_b^2\right)+\vec{p}_a^2\vec{p}_b^2}{\sqrt{m^2+\vec{p}_a^2}\sqrt{m^2+\vec{p}_b^2}+m^2}. \qquad (11)$$

Після цього матимемо

$$\triangle D = 2\sum_{a=1}^{k-1}\sum_{b=a+1}^{k}\left[\frac{1}{\sqrt{\left(m^2+\vec{p}_a^2\right)\left(m^2+\vec{p}_b^2\right)}+m^2}\right.$$
$$\left.\times\left(m^2\left(\vec{p}_a^2+\vec{p}_b^2\right)+\vec{p}_a^2\vec{p}_b^2 - (\vec{p}_a\cdot\vec{p}_b)\left(\sqrt{m^2+\vec{p}_a^2}\sqrt{m^2+\vec{p}_b^2}+m^2\right)\right)\right]. \qquad (12)$$

В чисельнику цього виразу зручно перегрупувати доданки таким чином:

$$m^2\left(\vec{p}_a^2+\vec{p}_b^2\right)+\vec{p}_a^2\vec{p}_b^2 - (\vec{p}_a\cdot\vec{p}_b)\left(\sqrt{m^2+\vec{p}_a^2}\sqrt{m^2+\vec{p}_b^2}+m^2\right) =$$
$$= \frac{1}{2}m^2\vec{p}_a^2 + \frac{1}{2}\vec{p}_a^2\vec{p}_b^2 + \frac{1}{2}m^2\vec{p}_b^2 + \frac{1}{2}\vec{p}_a^2\vec{p}_b^2$$
$$- (\vec{p}_a\cdot\vec{p}_b)\sqrt{m^2+\vec{p}_a^2}\sqrt{m^2+\vec{p}_b^2}$$
$$+ \frac{1}{2}m^2\vec{p}_a^2 - m^2(\vec{p}_a\cdot\vec{p}_b) + \frac{1}{2}m^2\vec{p}_b^2. \qquad (13)$$

Після цього замість (12) легко отримати:

$$\triangle D = \sum_{a=1}^{k-1}\sum_{b=a+1}^{k} \frac{\left(\vec{p}_a\sqrt{m^2+\vec{p}_a^2} - \vec{p}_b\sqrt{m^2+\vec{p}_a^2}\right)^2 + m^2\left(\vec{p}_a - \vec{p}_b\right)^2}{\left(m^2+\vec{p}_a^2\right)\left(m^2+\vec{p}_b^2\right) + m^2}. \quad (14)$$

Як видно з виразів (14) і (6)

$$\triangle D \geq 0, \quad F = \frac{1}{\left(k^2-1\right)m^2 + \triangle D} \leq \frac{1}{\left(k^2-1\right)m^2}. \quad (15)$$

В той же час, з цих виразів видно, що, якщо хоча б якісь два з чотириімпульсів, які входять в суму в виразі (2), не дорівнюють один одному, то нерівності (15) стають строгими. Отже, кожний множник типу (6) досягає максимального значення лише за умови рівності між собою всіх чотириімпульсів, які в нього входять. Деревний блок є добутком множників типу (6). Оскільки кожен з цих множників є додатним, то нерівності (15) для всіх ліній деревного блоку можна перемножити між собою. При цьому за умови рівності між собою чотиривекторів енергії-імпульсу всіх реальних частинок, що є вихідними для цього блоку, в усіх нерівностях реалізуються саме рівності.

Звідси можна зробити висновок, що деревний блок досягає свого максимального значення за умови рівності між собою чотиривекторів енергії-імпульсу всіх реальних частинок, які з цього блоку виходять. Тобто такий деревний блок буде описувати народження частинок зі скорельованими чотириімпульсами і тому його можна розглядати як модель адронної зливи [12].

З наведених міркувань випливає, що умовою максимальності множника, яка відповідає кожному деревному блоку на рис.1, є рівність між собою тривимірних компонент імпульсів всіх вторинних частинок, що приєднуються до блоку. При цьому саме спільне значення імпульсів може бути яким завгодно, бо деревний блок від нього не залежить внаслідок умови масової поверхні для кожної вторинної частинки. Але ці значення імпульсів для кожного деревного блоку можуть бути встановлені шляхом максимізації модуля виразу, який відповідає усій діаграмі рис.1.

Таким чином, модель, заснована на діаграмах типу рис. 1, виявляється аналогічною моделі, розглянутої в роботі [1]. Аналогом груп частинок, в яких максимум досягається при однакових значеннях чотириімпульсів частинок в групі, виступають деревні блоки. До того ж, застосування методу Лапласа спрощує процедуру побудови блоків з урахуванням всіх можливих діаграм без петель. В межах методу Лапласа щодо кожного такого блоку нам достатньо знайти лише його значення в точці максимуму (тобто при рівних імпульсах вторинних частинок), а також другі похідні по компонентах цих імпульсів в тій самій точці максимуму.

## 4. Рекурентна процедура розрахунку значення деревного блока і його похідних в точці максимуму

Для дерев з двома і трьома лініями вторинних частинок, всі варіанти діаграм перелічено на рис. 3 - 4. Для них легко розрахувати і значення в точці максимуму, і другі похідні в точці максимуму. Розрахунок потрібних величин для дерев із більшою кількістю вторинних частинок можна провести рекурентно. Відповідна процедура пояснюється на рис.7.

Як видно з рис.7, вхідна лінія діаграми розгалужується на дві лінії, до кожної з яких приєднується дерево з меншою кількістю вторинних частинок. При цьому під $k$ і $n-k$ на рис.7 ми розуміємо саме суму всіх можливих безпетльових діаграм із відповідною кількістю ліній вторинних частинок. Вираз, що зіставляється деревам $k$ і $n-k$, можна побудувати, рухаючись уздовж діаграми від вихідних ліній до вхідної. Тому вирази для блоків $k$ і $n-k$ будуть однаковими як у випадку, коли вони є «самостійними» деревами, що приєднуються до діаграми типу рис.1, так і у випадку, коли вони приєднуються до гілок на рис.7.

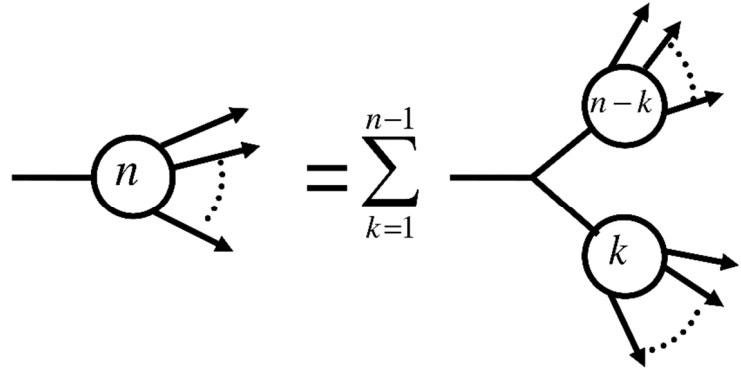

Рис. 7. Рекурентна процедура побудови деревного блоку з $n$ лініями вторинних частинок.

Записуючи аналітичний вираз, що відповідає діаграмі рис. 7, і покладаючи імпульси частинок рівними між собою, можемо знайти величину блоку $n$ в точці максимуму через величини, розраховані на попередніх кроках рекурентної процедури. Так само, диференціюючи вираз, що відповідає рис. 7, потрібну кількість разів можна знайти похідні довільного порядку від блоку $n$ в точці максимуму через величини, розраховані на попередніх кроках.

Розглянемо докладніше процедуру розрахунку похідних від деревного блоку типу, зображеного на рис.7. Позначимо аналітичний вираз для цього блоку як $B_n(p_1, p_2, ..., p_n)$. Аргументами цієї функції є чотириімпульси вторинних частинок, яким відповідають лінії, що приєднуються до блоку. Оскільки значення функції $B_n(p_1, p_2, ..., p_n)$ є Лоренц-інваріантом, то і залежати вона повинна від Лоренц-інваріантів. Такими Лоренц-інваріантами є всі можливі скалярні добутки $(p_a p_b)$, $a = 1...n$, $b = a...n$. Позначаючи множину всіх цих добутків як $\{(p_a p_b)\}$, можемо записати $B_n(p_1, p_2, ..., p_n) = B_n(\{(p_a p_b)\})$. В точці максимуму дерева всі чотириімпульси дорівнюють один одному. Це не залежить від того, чи є це дерево самостійним блоком, як $B_n$ на рис. 7, чи воно є частиною більшого дерева, як $B_k$ або $B_{n-k}$ на тому ж рисунку. В точці максимуму таким чином будь-яка множина $\{(p_a p_b)\}$ зводиться до такої, що містить лише один елемент $\{m^2\}$. Тому значення будь якого дерева в точці максимуму однакове незалежно від розташування цього дерева на діаграмі. Наприклад, якщо деревний блок $B_k$ є частиною двох різних дерев, що приєднуються до різних вершин на діаграмі рис. 7, в точці максимуму він буде мати однакові значення.

Розглянемо тепер перші похідні від блоку $B_n(\{p_a p_b\})$ по компонентах тривимірних імпульсів частинок, які приєднуються до блоку. Ці компоненти позначатимемо $(\vec{p}_a)^b$, $a = 1\ldots n$, $b = 1\ldots 3$. Тут $a$ позначає номер імпульсу як аргументу функції $B_n(\{p_a p_b\})$, а $b$ – номер просторової компоненти цього імпульсу. Ці просторові компоненти імпульсів входять до виразу для $B_n(\{p_a p_b\})$ лише через компоненти чотириімпульсів частинок, які позначатимемо $(p_e)^f$, $e = 1\ldots n$, $f = 0\ldots 3$. З урахуванням цього маємо

$$\frac{\partial B_n(\{p_a p_b\})}{\partial (\vec{p}_c)^d} = \sum_{l=0}^{3} \frac{\partial B_n(\{p_a p_b\})}{\partial (p_c)^l} \frac{\partial (p_c)^l}{\partial (\vec{p}_c)^d} \qquad (16)$$

Похідні $\partial (p_c)^l / \partial (\vec{p}_c)^d$ легко розрахувати явно:

$$\frac{\partial (p_c)^l}{\partial (\vec{p}_c)^d} = \frac{(\vec{p}_c)^d}{\sqrt{m^2 + \vec{p}_c^{\,2}}} \delta_0^l + \delta_d^l. \qquad (17)$$

Внаслідок цього їх розрахунок легко реалізувати при чисельному розрахунку у вигляді відповідної підпрограми.

Розглянемо четвірку похідних $\dfrac{\partial B_n(\{p_a p_b\})}{\partial (p_c)^l}$, $l = 0\ldots 3$. Оскільки мова йде про похідні від скалярної функції по компонентам контраваріантного чотири-вектора, то вони утворюють коваріантний чотири-вектор. Оскільки єдині чотири-вектори, що входять у $B_n(\{p_a p_b\})$, є лише чотири-вектори енергії-імпульсу частинок, що приєднуються до дерева, то й похідні можуть залежати лише від них. Тому можемо написати $\dfrac{\partial B_n(\{p_a p_b\})}{\partial (p_c)^l}$ як

$$\frac{\partial B_n(\{p_a p_b\})}{\partial (p_c)^l} = \sum_{d=1}^{n} K_{ndc}(\{p_a p_b\})(p_d)_l. \qquad (18)$$

Тут $K_{ndc}(\{p_a p_b\})$ - коефіцієнтні функції. Їх вирази через їх аргументи залежать від внутрішньої структури дерева $B_n$, а значення визначаються скалярними добутками $\{p_a p_b\}$.

Формулу (18) можна обґрунтувати ще й таким чином. Всі чотириімпульси входять в вираз для блоку $B_n$ лише через Лоренц-інваріанти, що є квадратичними по їх компонентах. Тоді диференціювання по компоненті чотири-імпульсу передбачатиме суму добутків похідних від $B_n$ по відповідних Лоренц-інваріантах на похідні від цих Лоренц-інваріантів по компонентах чотириімпульсів. Перші з цих множників самі є Лоренц-інваріантами. Другі, як похідні від квадратичних виразів, є лінійними по компонентах чотириімпульсів. Тоді, збираючи коефіцієнти при кожній компоненті, отримаємо вираз (18).

Оскільки права частина рівності (18) є коваріантним чотиривектором, а ліва – лінійною комбінацією коваріантних чотиривекторів, то дістаємо висновку, що коефіцієнти $K_{ndc}(\{p_a p_b\})$ повинні бути скалярами, тобто маємо скалярні функції від множини $\{p_a p_b\}$. Тому кожна з функції $K_{ndc}(\{p_a p_b\})$ аналогічна за своїми властивостями функції $B_n(\{p_a p_b\})$. Це означає, що значення коефіцієнтів $K_{ndc}(\{p_a p_b\})$ в точці максимуму залежать лише від виду дерева, але не залежать від того, як це дерево приєднане до діаграми. Тому в точці максимуму, де всі чотириімпульси однакові, маємо

$$\frac{\partial B_n(\{p_a p_b\})}{\partial (p_c)^l} = \left[\sum_{d=1}^{n} K_{ndc}(m^2)\right] p_l. \qquad (19)$$

При цьому коефіцієнт $K_{ndc}(m^2)$ не залежить від спільного значення всіх чотириімпульсів. Відтак, він не залежить від того, як саме дерево приєднується до діаграми. Скористаймося цим результатом, щоб побудувати перші похідні від деревних блоків $B_n(\{p_a p_b\})$. Задля цього зробимо декілька зауважень щодо процедури рекурентної побудови деревних блоків.

Типова вершина теорії «фі-три» показана на рис. 8.

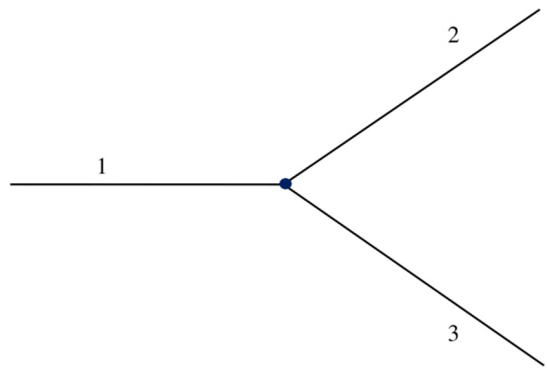

Рис. 8. Типова вершина теорії «фі-три». Щоб підкреслити індивідуальність кожної лінії вони були пронумеровані.

Нагадаємо, що три лінії, які виходять з вершини, відповідають трьом операторним множникам, що входять в лагранжіан взаємодії. Це є три різні множники, які дорівнюють один одному. Це означає, що, якщо ми з одним і тим самим оператором, який входить або до іншої вершини, або відповідає зовнішній лінії, спаримо один раз оператор, який відповідає лінії 2 на рис. 8, а інший - лінії 3, то ми отримаємо різні варіанти спарювань. Згідно з теоремою Віка, обидва ці варіанти спарювань повинні бути враховані, тобто, якщо на будь-якій діаграмі ми переставимо 3! способами всі лінії, що приєднуються до кожної вершини, то отримаємо різні доданки в сумі по всіх спарюваннях, які повинні бути складені за теоремою Віка.

З іншого боку, всі ці доданки будуть рівними. Їх кількість дорівнюватиме $(3!)^{n_v}$ де $n_v$ – кількість вершин діаграми. Тому при побудові діаграм ми можемо не враховувати діаграми, які отримуються одна з одної перестановкою ліній, що відповідають одній й тій самій вершині, але замість цього помножити внесок цієї діаграми на $(3!)^{n_v}$. Оскільки таким же чином в вираз для внеску діаграми в амплітуду розсіяння входитиме й константа зв'язку $g$, то можна ввести нову константу зв'язку $g' = 3!g$ і, тим самим, врахувати внески від діаграм, що відрізняються перестановкою ліній однієї й тієї ж вершини.

Приймаючи таке правило побудови діаграм, розглянемо процедуру рекурентної побудови блоку $B_n(\{p_a p_b\})$. Припустимо, що в нас вже побудовані деревні блоки, які відповідають меншій ніж $n$ кількості вторинних частинок. Якщо $k_1 < n$, $k_2 < n$ та $k_1 + k_2 = n$, то процес побудови блоку $B_n$ з блоків $B_{k_1}$ і $B_{k_2}$ пояснюється на рис. 9.

Лінію 1 залишимо для приєднання до решти діаграми рис. 1. Блок $B_{k_1}$ приєднаємо до лінії 2, а блок $B_{k_2}$ – до лінії 3. Якщо ми приєднаємо $B_{k_1}$ до лінії 3, а $B_{k_2}$ – до лінії 2, то отримаємо діаграму, яка відрізняється від попередньої лише перестановкою ліній 2 і 3. Тому такі діаграми не повинні враховуватися як різні за обговореним вище правилом. Також не повинні враховуватися діаграми, які відрізняються між собою лише перестановкою $B_{k_1}$ і $B_{k_2}$. Щоб досягти цього приймемо правило: якщо $k_1 \neq k_2$, то дерево з більшою кількістю вторинних частинок приєднується до лінії 2 на рис. 9, а з меншою - до лінії 3. У випадку коли $k_1 = k_2$, діаграму будемо враховувати лише один раз, не переставляючи ці блоки між собою.

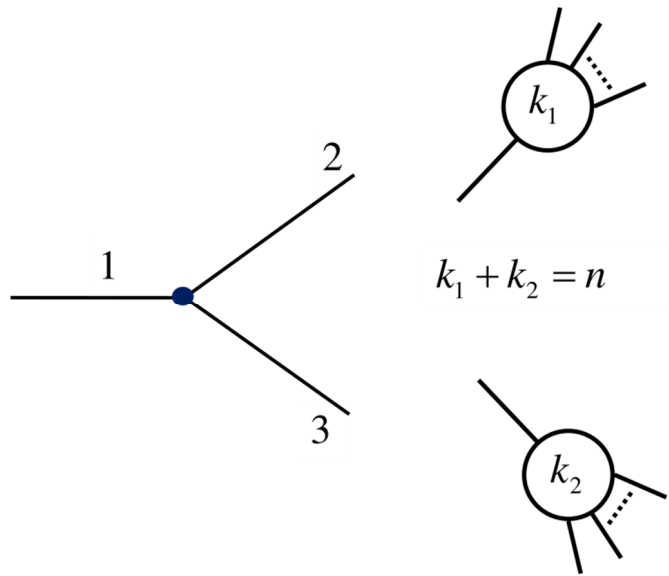

Рис. 9. Процес побудови деревного блоку з інших блоків з меншою кількістю вторинних частинок. Відсутність стрілок у вихідних ліній підкреслює, що на даному рисунку зображено неспарені лінії, які потім всіма можливими способами повинні бути спарені із лініями вторинних частинок.

При цьому множник, який відповідає вхідній лінії, включаємо до блоку, як це показано на рис. 10. Окрім того, на цьому етапі обиратимемо якийсь один спосіб приєднання зовнішніх ліній до діаграми, а суму по всіх перестановках врахуємо пізніше.

Враховуючи міркування, що наведені до формул (1) - (4), вираз, який відповідає блоку, зображеному на рис. 10, можемо записати:

$$B_n\left(\left\{p_a p_b \mid a,b = 1\ldots n\right\}\right) =$$
$$= \frac{1}{\left(\sum_{a=1}^{n} p_a\right)^2 - m^2} \sum_{k_2=1}^{[n/2]} \begin{bmatrix} B_{n-k_2}\left(\left\{p_a p_b \mid a,b = k_2+1\ldots n\right\}\right) \\ \times B_{k_2}\left(\left\{p_a p_b \mid a,b = 1\ldots k_2\right\}\right) \end{bmatrix}. \quad (20)$$

Тут через $[n/2]$ у верхній границі суми, як зазвичай, позначена ціла частина числа $n/2$. Позначення $B_n\left(\left\{p_a p_b \mid a,b = 1\ldots n\right\}\right)$ вказує на множину, яку утворюють всі можливі індекси чотириімпульсів, що утворюють скалярні добутки, які входять в множину $\{p_a p_b\}$. Розглянемо перших декілька кроків реалізації рекурентної процедури, заснованої на співвідношенні (20).

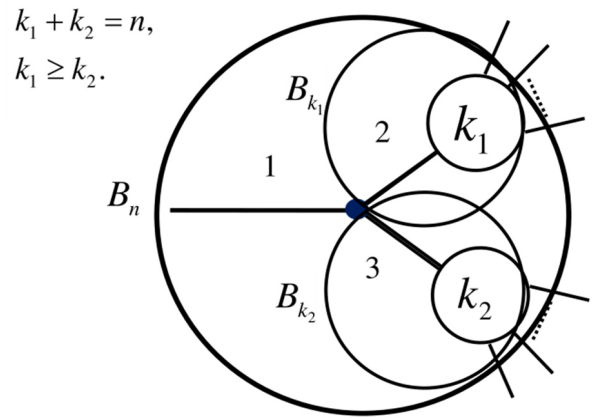

Рис. 10. Схематичне зображення блоків, що ілюструє включення вхідної лінії до блоку.

Блок $B_1$ містить лише одну зовнішню лінію і за правилами діаграмної техніки цій лінії нічого не зіставляється, тому

$$B_1 = 1, \quad \frac{\partial B_1}{\partial (p_c)^l} = 0. \qquad (21)$$

Блок $B_2\left(\left\{p_1^2, p_1 p_2, p_2^2\right\}\right)$ зображено на рис. 3 та його аналітичний вираз має вигляд

$$B_2\left(\left\{p_1^2, p_1 p_2, p_2^2\right\}\right) = \frac{1}{(p_1 + p_2)^2 - m^2}. \qquad (22)$$

Значення цього блоку в точці максимуму $p_1 = p_2 = p$ (де $p$ – спільне значення чотиріімпульсів $p_1$ і $p_2$ в точці максимуму, яке ми отримуємо шляхом максимізації всього внеску від діаграми рис. 1 в амплітуду розсіяння)

$$B_2^{\max} = \frac{1}{3m^2}. \qquad (23)$$

Щодо похідних від цього блоку, то нагадаємо, що згідно з формулою (16), для розрахунку похідних по компонентах тривимірних імпульсів, достатньо розрахувати похідні по компонентах чотириімпульсів. Їх подальша згортка (16) з величинами (17), які легко обчислити по відомих значеннях імпульсів в точці максимуму, дасть нам потрібні величини. Для блоку $B_2$ розрахунок цих похідних пояснюється наступною послідовністю перетворень:

$$\frac{\partial B_2\left(\left\{p_1^2, p_1 p_2, p_2^2\right\}\right)}{\partial (p_1)^c} = \frac{\partial}{\partial (p_1)^c} \frac{1}{(p_1+p_2)^2 - m^2} =$$
$$= \frac{\partial}{\partial (p_1)^c} \frac{1}{g_{ab}(p_1+p_2)^a (p_1+p_2)^b - m^2} = \qquad (24)$$
$$= -\frac{1}{(p_1+p_2)^2 - m^2} \frac{\partial}{\partial (p_1)^c}\left(g_{ab}(p_1+p_2)^a (p_1+p_2)^b\right) =$$
$$= -\frac{2}{\left[(p_1+p_2)^2 - m^2\right]^2}\left[(p_1)_c + (p_2)_c\right].$$

Аналогічно, отримаємо й похідні по компонентах чотиривектора $p_2$:

$$\frac{\partial B_2\left(\left\{p_1^2, p_1 p_2, p_2^2\right\}\right)}{\partial (p_2)^c} = -\frac{2}{\left[(p_1+p_2)^2 - m^2\right]^2}\left[(p_1)_c + (p_2)_c\right] \qquad (25)$$

Виходячи з цього та (18), маємо

$$K_{211} = K_{221} = K_{212} = K_{222} = -\frac{2}{\left[(p_1+p_2)^2 - m^2\right]^2} \qquad (26)$$

при рівних чотириімпульсах в точці максимуму отримаємо

$$K_{211} = K_{221} = K_{212} = K_{222} = -\frac{2}{9m^4} \qquad (27)$$

Отже, як зазначено вище, ці коефіцієнти в точці максимуму не залежать від чотириімпульсів і можуть бути розраховані для блоку $B_2$ незалежно від того, як він підключений до діаграми. Вводячи позначення

$$K_{21} = -\frac{4}{9m^4}, \ K_{22} = -\frac{4}{9m^4}, \qquad (28)$$

величини похідних в точці максимуму можемо записати:

$$\left(\frac{\partial B_2}{\partial (p_1)^c}\right)^{\max} = K_{21} p_c, \ \left(\frac{\partial B_2}{\partial (p_2)^c}\right)^{\max} = K_{22} p_c. \qquad (29)$$

Для блоку $B_3(\{p_a p_b\})$, у відповідності із розглянутими вище правилами, маємо одну діаграму, яку наведено на рис. 11:

$$B_3(p_1, p_2, p_3) = \frac{1}{(p_1 + p_1 + p_3)^2 - m^2} B_2(p_2, p_3). \qquad (30)$$

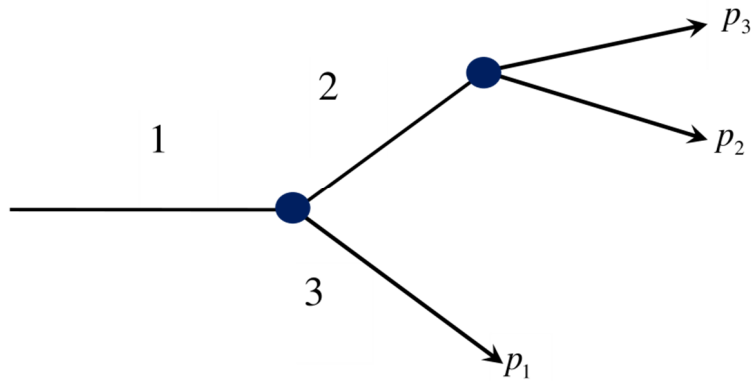

Рис. 11. Діаграма блоку $B_3(\{p_a p_b\})$.

Диференціюючи співвідношення (30), маємо

$$\frac{\partial B_3(p_1, p_2, p_3)}{\partial (p_1)^a} = -\frac{2(p_1 + p_1 + p_3)_a}{\left[(p_1 + p_1 + p_3)^2 - m^2\right]^2} B_2(p_2, p_3), \qquad (31)$$

$$\frac{\partial B_3(p_1,p_2,p_3)}{\partial (p_2)^a} = -\frac{2(p_1+p_1+p_3)_a}{\left[(p_1+p_1+p_3)^2-m^2\right]^2}B_2(p_2,p_3)$$
$$+\frac{1}{(p_1+p_1+p_3)^2-m^2}\frac{\partial B_2(p_2,p_3)}{\partial (p_2)^a},$$
(32)

$$\frac{\partial B_3(p_1,p_2,p_3)}{\partial (p_3)^a} = -\frac{2(p_1+p_1+p_3)_a}{\left[(p_1+p_1+p_3)^2-m^2\right]^2}B_2(p_2,p_3)$$
$$+\frac{1}{(p_1+p_1+p_3)^2-m^2}\frac{\partial B_2(p_2,p_3)}{\partial (p_3)^a}.$$
(33)

В точці максимуму, тобто при $p_1 = p_1 = p_3 = p$, де компоненти чотиривектора $p$ визначаються тим, до якої вершини на рис. 1 приєднується дерево $B_3$, або те дерево, частиною якого воно є, маємо:

$$B_3^{max} = \frac{B_2^{max}}{8m^2}$$
$$\left(\frac{\partial B_3}{\partial (p_1)^a}\right)^{max} = \left(-\frac{3B_2^{max}}{32m^4}\right)p_a$$
$$\left(\frac{\partial B_3}{\partial (p_2)^a}\right)^{max} = \left(\frac{K_{21}}{8m^2}-\frac{3B_2^{max}}{32m^4}\right)p_a$$
$$\left(\frac{\partial B_3}{\partial (p_2)^a}\right)^{max} = \left(\frac{K_{22}}{8m^2}-\frac{3B_2^{max}}{32m^4}\right)p_a.$$
(34)

Величини $B_2^{\max}$ і $K_{21}$, $K_{22}$, що входять до цих співвідношень, визначаються формулами (26) і (28), тобто вони відомі з попередніх кроків розглядуваної рекурентної процедури.

Для подальшого введемо такі позначення

$$\left(\frac{\partial B_3}{\partial (p_i)^a}\right)^{max} = K_{3i} p_a, \ i=1,2,3 \qquad (35)$$

$$K_{31} = -\frac{3B_2^{max}}{32m^4}, K_{32} = K_{33} = \frac{K_{22}}{8m^2} - \frac{3B_2^{max}}{32m^4}.$$

Отже будемо й далі записувати похідні від блоків в точці максимуму в вигляді

$$\left(\frac{\partial B_c}{\partial (p_d)^a}\right)^{max} = K_{cd} p_a. \qquad (36)$$

Розглянемо тепер перші похідні від блоку $B_n$

$$\frac{\partial}{\partial (p_d)^l} B_n\left(\{p_a p_b \mid a,b=1...n\}\right) =$$

$$= \frac{\partial}{\partial (p_d)^l} \frac{1}{\left(\sum_{a=1}^n p_a\right)^2 - m^2} \sum_{k_2=1}^{[n/2]} \left[\begin{matrix} B_{n-k_2}\left(\{p_a p_b \mid a,b=k_2+1...n\}\right) \\ \times B_{k_2}\left(\{p_a p_b \mid a,b=1...k_2\}\right) \end{matrix}\right] \qquad (37)$$

$$+ \frac{1}{\left(\sum_{a=1}^n p_a\right)^2 - m^2} \sum_{k_2=1}^{[n/2]} \frac{\partial}{\partial (p_d)^l} \left[\begin{matrix} B_{n-k_2}\left(\{p_a p_b \mid a,b=k_2+1...n\}\right) \\ \times B_{k_2}\left(\{p_a p_b \mid a,b=1...k_2\}\right) \end{matrix}\right].$$

і розглянемо детально перший доданок виразу (37)

$$\frac{\partial}{\partial (p_d)^l} \frac{1}{\left(\sum_{a=1}^n p_a\right)^2 - m^2} \sum_{k_2=1}^{[n/2]} \left[\begin{matrix} B_{n-k_2}\left(\{p_a p_b \mid a,b=k_2+1...n\}\right) \\ \times B_{k_2}\left(\{p_a p_b \mid a,b=1...k_2\}\right) \end{matrix}\right] =$$

$$= -\frac{\left(\sum_{a=1}^n p_a\right)_l}{\left[\left(\sum_{a=1}^n p_a\right)^2 - m^2\right]^2} \sum_{k_2=1}^{[n/2]} \left[\begin{matrix} B_{n-k_2}\left(\{p_a p_b \mid a,b=k_2+1...n\}\right) \\ \times B_{k_2}\left(\{p_a p_b \mid a,b=1...k_2\}\right) \end{matrix}\right] \qquad (38)$$

Враховуючи вираз (20) для $B_n$, цей доданок можна переписати у вигляді

$$-\frac{\left(\sum_{a=1}^{n} p_a\right)_l}{\left[\left(\sum_{a=1}^{n} p_a\right)^2 - m^2\right]^2} \sum_{k_2=1}^{[n/2]} \left[\begin{array}{c} B_{n-k_2}\left(\{p_a p_b \mid a,b=k_2+1\ldots n\}\right) \\ \times B_{k_2}\left(\{p_a p_b \mid a,b=1\ldots k_2\}\right) \end{array}\right] =$$

$$= -\frac{\left(\sum_{a=1}^{n} p_a\right)_l}{\left(\sum_{a=1}^{n} p_a\right)^2 - m^2} B_n\left(\{p_a p_b \mid a,b=1\ldots n\}\right).$$

(39)

Тоді вираз (37) приймає вигляд

$$\frac{\partial}{\partial (p_d)^l} B_n\left(\{p_a p_b \mid a,b=1\ldots n\}\right) =$$

$$= -\frac{\left(\sum_{a=1}^{n} p_a\right)_l}{\left(\sum_{a=1}^{n} p_a\right)^2 - m^2} B_n\left(\{p_a p_b \mid a,b=1\ldots n\}\right)$$

$$+ \frac{1}{\left(\sum_{a=1}^{n} p_a\right)^2 - m^2} \sum_{k_2=1}^{[n/2]} \left[\begin{array}{c} \frac{\partial B_{n-k_2}}{\partial (p_d)^l}\left(\{p_a p_b \mid a,b=k_2+1\ldots n\}\right) \\ \times B_{k_2}\left(\{p_a p_b \mid a,b=1\ldots k_2\}\right) \\ + B_{n-k_2}\left(\{p_a p_b \mid a,b=k_2+1\ldots n\}\right) \\ \times \frac{\partial B_{k_2}}{\partial (p_d)^l}\left(\{p_a p_b \mid a,b=1\ldots k_2\}\right) \end{array}\right].$$

(40)

Відповідно в точці максимуму отримаємо

$$\left(\frac{\partial B_n}{\partial (p_d)^l}\right)^{max} = K_{nd} p_l,$$

(41)

де вираз для $K_{nd}$ легко отримується з (40)

$$K_{nd} = -\frac{1}{(n^2-1)m^2}\left[nB_n^{\max} + \sum_{k_2=1}^{[n/2]} \left(K_{n-k_2,d-k_2} B_{k_2}^{\max} + B_{n-k_2}^{\max} K_{k_2,d}\right)\right].$$

(42)

При цьому ми прийняли нумерацію зовнішніх імпульсів, за тим самим принципом, що й на рис. 11, тобто нумеруємо імпульси, починаючи від блоку з меншою кількістю частинок. Тому блок $B_{n-k_2}$ залежіть від імпульсів з індексами, починаючи від $k_2+1$. Тому в формулі (42) треба покласти

$$\begin{cases} K_{n-k_2, d-k_2} = 0, & \text{при } d - k_2 < 1 \\ K_{k_2, d} = 0, & \text{при } d > k_2 \end{cases}. \tag{43}$$

Тепер розглянемо другі похідні від $B_n$, необхідні для застосування методу Лапласа та описаного в попередніх розділах методу врахування інтерференційних внесків. Диференціюючи співвідношення (16), отримаємо

$$\frac{\partial^2 B_n(\{p_a p_b\})}{\partial (\vec{p}_e)^g \partial (\vec{p}_c)^d} = \sum_{l=0}^{3} \left( \sum_{h=0}^{3} \frac{\partial B_n(\{p_a p_b\})}{\partial (p_e)^h \partial (p_c)^l} \frac{\partial (p_e)^h}{\partial (\vec{p}_e)^g} \right) \frac{\partial (p_c)^l}{\partial (\vec{p}_c)^d} \\ + \sum_{l=0}^{3} \frac{\partial B_n(\{p_a p_b\})}{\partial (p_c)^l} \frac{\partial^2 (p_c)^l}{\partial (\vec{p}_e)^g \partial (\vec{p}_c)^d} \tag{44}$$

Враховуючи, що компоненти чотириімпульсу $p_c$ залежать лише від компонент тривимірного імпульсу з тим самим індексом, маємо

$$p_c = \begin{pmatrix} \sqrt{m^2 + \vec{p}_c^{\,2}} \\ (\vec{p}_c)^1 \\ (\vec{p}_c)^2 \\ (\vec{p}_c)^3 \end{pmatrix}, \tag{45}$$

можемо записати

$$\frac{\partial^2 (p_c)^l}{\partial (\vec{p}_e)^g \partial (\vec{p}_c)^d} = \delta_{ec} \frac{\partial^2 (p_c)^l}{\partial (\vec{p}_c)^g \partial (\vec{p}_c)^d}, \tag{46}$$

де $\delta_{ec}$ – символ Кронекера.

З урахуванням (17) отримаємо

$$\frac{\partial^2 (p_c)^l}{\partial (\vec{p}_c)^g \partial (\vec{p}_c)^d} = \frac{\delta_0^l}{\sqrt{m^2 + \vec{p}_c^2}} \left( \delta_{gd} - \frac{(\vec{p}_c)^d (\vec{p}_c)^g}{m^2 + \vec{p}_c^2} \right) \tag{47}$$

В точці максимуму, коли всі чотиривектори однакові, маємо

$$\left( \frac{\partial^2 B_n}{\partial (\vec{p}_e)^g \partial (\vec{p}_c)^d} \right)^{\max} = \sum_{l=0}^{3} \sum_{h=0}^{3} \left( \frac{\partial B_n}{\partial (p_e)^h \partial (p_c)^l} \right)^{\max} \frac{\partial p^h}{\partial (\vec{p})^g} \frac{\partial p^l}{\partial (\vec{p})^d}$$
$$+ \sum_{l=0}^{3} \frac{\partial B_n}{\partial (p_c)^l} \delta_{ec} \left( \delta_{gd} - \frac{(\vec{p})^d (\vec{p})^g}{m^2 + \vec{p}^2} \right) \frac{\delta_0^l}{\sqrt{m^2 + \vec{p}^2}}. \tag{4.48}$$

Щодо цього виразу, то залишається розрахувати компоненти тензору $\left( \frac{\partial B_n}{\partial (p_e)^h \partial (p_c)^l} \right)^{\max}$. Компоненти цього тензору повинні виражатися через компоненти спільного значення всіх чотириімпульсів дерева в точці максимуму $p_a$. При цьому оскільки тензор двічі коваріантний, він повинен виражатися через коваріантні компоненти чотиривектору. З цих компонент можна побудувати лише один двічі коваріантний тензор $p_a p_b$. Цей тензор, як і тензор других похідних, є симетричним. Окрім того, компоненти довільного симетричного тензору можуть виражатися через компоненти метричного тензору $g_{ab}$. Тому загальна структура тензору $\left( \frac{\partial B_n}{\partial (p_e)^h \partial (p_c)^l} \right)^{\max}$ повинна бути наступною

$$\left( \frac{\partial B_n}{\partial (p_e)^h \partial (p_c)^l} \right)^{\max} = D_{nec}^{(1)} g_{hl} + D_{nec}^{(2)} p_h p_l. \tag{48}$$

Коефіцієнти $D_{nec}^{(1)}$ і $D_{nec}^{(2)}$ будуть розраховано рекурентним способом. Розглянемо цю процедуру на прикладі, як і для перших похідних. Оскільки блок

$B_1$ є константою, всі його похідні дорівнюють нулю. Далі розглянемо блок $B_2$. Диференціюючи співвідношення (25), отримаємо:

$$\frac{\partial^2 B_2\left(\left\{p_1^2, p_1 p_2, p_2^2\right\}\right)}{\partial(p_a)^d \partial(p_b)^c} = \frac{8}{\left[(p_1+p_2)^2 - m^2\right]^3}\left[(p_1)_d + (p_2)_d\right] \times \left[(p_1)_c + (p_2)_c\right] - \frac{2}{\left[(p_1+p_2)^2 - m^2\right]^2} g_{dc}, \quad (49)$$

де $a = 1,2$ та $b = 1,2$.

В цьому співвідношенні передбачається, що індекси $a$ і $b$ пробігають чотири пари значень $(1,1)$, $(1,2)$, $(2,1)$, $(2,2)$ і для всіх цих пар результат подвійного диференціювання виявляється однаковим. В точці максимуму маємо:

$$\left(\frac{\partial^2 B_2}{\partial(p_b)^c \partial(p_b)^c}\right)^{\max} = -\frac{2}{9m^4} g_{dc} + \frac{32}{27 m^6} p_d p_c. \quad (50)$$

Отже, порівнюючи з (48), отримуємо:

$$D_{2ab}^{(1)} = -\frac{2}{9m^4}, \quad D_{2ab}^{(2)} = \frac{32}{27 m^6}. \quad (51)$$

Вирази для коефіцієнтів $D_{nec}^{(1)}$ та $D_{nec}^{(2)}$ у виразі (48) можна знайти, якщо зібрати доданки з $p_h p_l$ та $g_{hl}$ після диференціювання виразу (40) (процедуру диференціювання цього виразу детально розглянуто у додатку):

$$D_{ncd}^{(1)} = \frac{1}{(n^2-1)m^2}\left[\sum_{k_2=1}^{[n/2]}\left(B_{k_2}^{\max} D_{n-k_2,cd}^{(1)} + B_{n-k_2}^{\max} D_{k_2 cd}^{(1)}\right) - B_n^{\max}\right]. \quad (52)$$

$$D_{ncd}^{(2)} = \frac{1}{(n^2-1)m^2} \left[ \sum_{k_2=1}^{[n/2]} \left( B_{k_2}^{\max} D_{n-k_2,cd}^{(2)} + B_{n-k_2}^{\max} D_{k_2 cd}^{(2)} \right) \right.$$

$$+ \sum_{k_2=1}^{[n/2]} \left( K_{n-k_2,d-k_2} K_{k_2,c} + K_{n-k_2,c-k_2} K_{k_2,d} \right)$$

$$\left. - \sum_{k_2=1}^{[n/2]} \frac{1}{(n^2-1)m^2} \left( B_{k_2}^{\max} K_{n-k_2,d-k_2} + B_{n-k_2}^{\max} K_{k_2,d} \right) n \right]$$

$$+ \frac{1}{(n^2-1)m^2} n^2 B_n^{\max} - n K_{nc}. \qquad (53)$$

Запропонована рекурентна процедура розрахунку перших та других похідних від виразу $B_n$ дозволяє застосувати цей метод врахування інтерференційних внесків для розрахунку внесків у перерізи не тільки від діаграм типу «гребінка», а й від усіх можливих безпетльових діаграм.

Виявляється, що вираз $B_n$, який відповідає діаграмі виду рис. 1 з певною кількістю дерев $k$, має аналогічні властивості, що й вираз $A_n(\sqrt{s}, X)$ в роботі [1] для діаграми, вершини якої згруповані у $k$ груп: в точці максимуму виразу, що відповідає діаграмі виду рис. 1, імпульси вторинних частинок приймають однакові значення в середині кожного окремого дерева, так само як імпульси вторинних частинок в середині груп діаграми типу «гребінка» у точці максимуму функції $A_n(\sqrt{s}, X)$. Іншими словами, імпульси вторинних частинок, що приєднуються до одного дерева, скорельовані та приймають однакові значення в точці максимуму виразу $B_n$. Таким чином, моделювання процесів розсіяння з врахуванням діаграм виду рис. 1 дозволяє моделювати адронні і партонні зливи (каскади вторинних частинок зі скорельованими імпульсами).

# 5. Аналіз внесків діаграм різної топології в парціальний переріз розсіяння з утворенням n вторинних частинок

Згідно із теоремою Віка, ми повинні розрахувати суму діаграм типу рис. 1 з усіма можливими перестановками зовнішніх ліній між собою. При цьому, аналогічно [1] можна в цій сумі по перестановках спочатку згрупувати доданки, що відповідають перестановкам частинок всередині кожного деревного блока. Далі можна розглядати суму отриманих, таким чином, виразів по всіх перестановках частинок між деревними блоками на рис.1. Розглянемо суму по перестановках частинок всередині кожного деревного блока. Оскільки кожен доданок приймає максимальне значення при рівних чотириімпульсах частинок, то всі доданки в такій сумі будуть приймати максимальне значення при одних і тих самих значеннях аргументів. Якщо позначити як $n_t$ кількість частинок в деревному блоці, то значення суми в точці максимуму буде в $n_t!$ більшим ніж значення кожного доданку в точці максимуму. Тому, як зазначалося у вступі, на відміну від моделі ДГЛАП, врахування інтерференції всередині деревного блока призводить до факторіального підсилення константи зв'язку.

У цьому підрозділі наведено результати розрахунків парціальних перерізів для діаграм виду рис. 1 при різних значеннях енергії та кількості вторинних частинок. У якості мас вторинних частинок для розрахунків приймалася маса піона 0.135ГеВ. У якості мас частинок з чотириімпульсами $P_1, P_2, P_3, P_4$ приймалася маса протона. Отримані результати для порівняння наведено разом з відповідними розрахунками для діаграм типу «гребінка» розглянуто в роботі [1].

Таким чином, можна порівняти внески від безпетльових діаграм з різною топологією. На вертикальних осях на рис. 12 - 14 показано значення безрозмірної величини $\tilde{\sigma}_n = \sigma_n / L^n$, де $L$ – ефективна константа зв'язку [1]. З отриманих результатів видно, що деревні діаграми виду рис. 1 в певному діапазоні енергій забезпечують суттєвий внесок у повний переріз розсіяння. На рис.12 видно, що внесок у парціальний переріз від деревних діаграм виду рис.1 з одним, двома та трьома вихідними деревами значно перевищують відповідний внесок від діаграм типу «гребінка» у діапазоні енергій від порогової енергії непружного розсіяння до енергії $\sqrt{s} \approx 55$ ГеВ.

При цьому, з ростом кількості вторинних частинок $n$, ширина області, в якій деревні діаграми забезпечують більший внесок ніж діаграми типу «гребінка», зростає, як це видно з рис. 13 – 14.

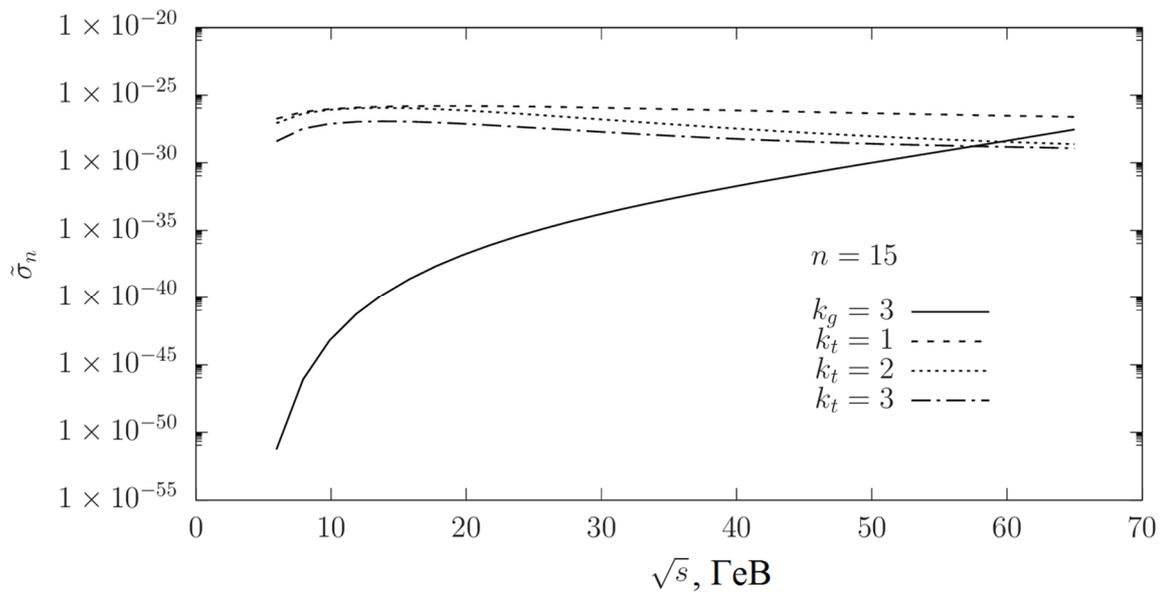

Рис. 12. Залежність внесків у парціальний переріз для діаграм виду «гребінка» (суцільна лінія) та деревних діаграм виду рис. 1 з одним, двома та трьома деревами (різні значення $k_t$) з утворенням $n = 15$ вторинних частинок.

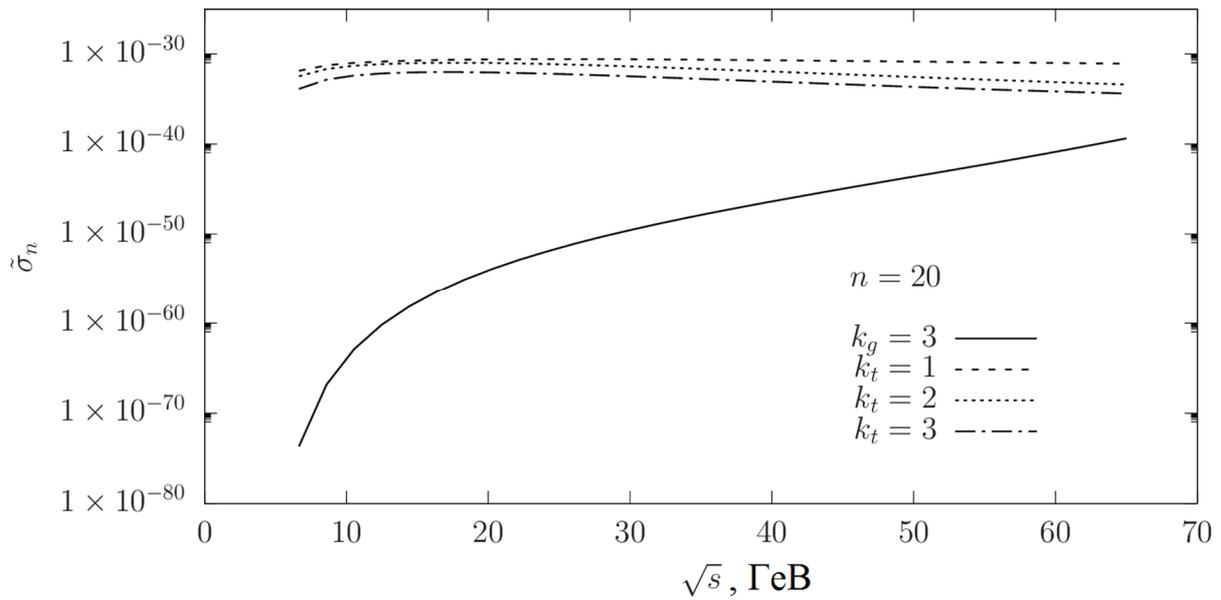

Рис. 13. Залежність внесків у парціальний переріз для діаграм виду «гребінка» (суцільна лінія) та деревних діаграм виду рис. 4.1 з одним, двома та трьома деревами (різні значення $k_t$) з утворенням $n = 20$ вторинних частинок.

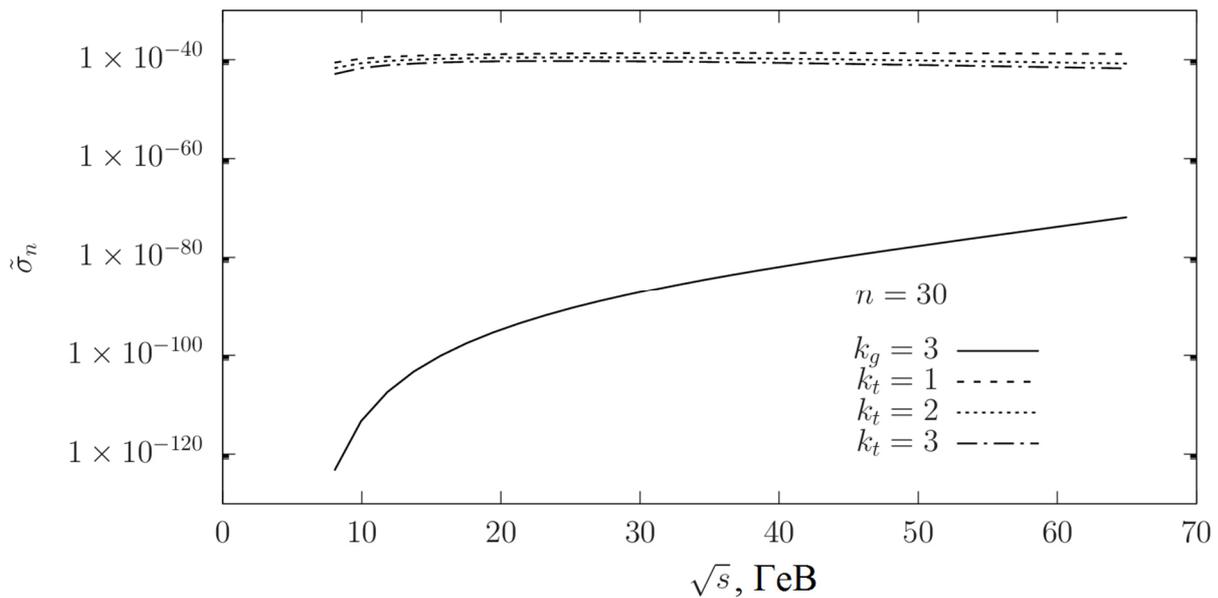

Рис. 14. Залежність внесків у парціальний переріз для діаграм виду «гребінка» (суцільна лінія) та деревних діаграм виду рис. 4.1 з одним, двома та трьома деревами (різні значення $k_t$) з утворенням $n = 30$ вторинних частинок.

Таким чином, отримані результати свідчать про необхідність врахування деревних діаграм виду рис. 1 при розрахунках перерізів та моделюванні процесів розсіяння. Крім того, як буде показано далі, при розгляді діаграм такого виду також виникає необхідність врахування інтерференційних внесків.

## 6. Необхідність врахування інтерференції між зливами

Як вже було показано вище, імпульси вторинних частинок, які приєднуються до одного дерева діаграми виду рис.1, приймають однакові значення в точці максимуму виразу, що відповідає цій діаграмі. Із міркувань симетрії випливає, що для діаграми з одним вихідним деревом $k_t = 1$, усі імпульси вторинних частинок будуть дорівнювати нулю, аналогічно, як це було для діаграми з однією групою вершин $k_g = 1$ в [1]. Зауважимо, що у випадку з групами, така властивість точки максимуму функції $A_n(\sqrt{s}, X)$ проявлялась вже після врахування перестановок вторинних частинок в середині груп, в той час як у випадку з деревами така властивість точки максимуму функції $B_n(\{p_a p_b\})$ проявляється ще до врахування інтерференції. В результаті чого для деревної діаграми з одним вихідним деревом будь-яка перестановка вторинних частинок не змінює точку максимуму функції. Тому при врахуванні інтерференційних внесків отримаємо суму великої кількості доданків, які приймають максимум в одній й тій самій точці.

Таким чином, відповідна функція $A_n(\sqrt{s}, X)$ (сума по всіх перестановках вторинних частинок в середині дерева, до якого вони приєднуються) для деревних діаграм завжди матиме єдину точку максимуму. Більш того, у випадку діаграм типу «гребінка» доданки, що утворювали суму $A_n(\sqrt{s}, X)$, хоч і мали *близькі* точки максимуму, але все ж відстань між точками максимуму цих доданків зростала з ростом енергії. Це призводило до того, що для фіксованих $n$ і $k_g$ з ростом енергії єдиний максимум функції $A_n(\sqrt{s}, X)$ ставав менш чітким (зменшувались величини власних значень матриці других похідних в точці максимуму) і функція починала набувати максимального значення в декількох точках. У випадку ж деревних діаграм, як вже було сказано, доданки суми $A_n\left(\sqrt{s}, X\right)$ матимуть єдину точку максимуму. Це призводить до того, що єдиний максимум функції $A_n\left(\sqrt{s}, X\right)$ не перетворюється у декілька окремих максимумів та не стає менш чітким внаслідок збільшення відстані між точками максимумів доданків суми $A_n\left(\sqrt{s}, X\right)$ з ростом енергії $\sqrt{s}$.

При цьому очікується, що з ростом енергії, максимуми функцій $A_n\left(\sqrt{s}, X\right)$ віддалятимуться один від одного, що призводитиме до того, що з ростом енергії інтерференційні внески, які відповідають перестановкам між різними деревами ставатимуть менш суттєвими. Усе це дозволяє ефективно застосовувати запропонований в [1] метод врахування інтерференційних внесків, заснований на методі Лапласа.

Розглянемо результати розрахунків внесків у парціальні перерізи розсіяння від діаграм виду рис.1 з урахуванням і без урахування інтерференційних внесків, що відповідають перестановкам між різними деревами. При цьому по аналогії з роботою [1] при розрахунках з декількома деревами вторинні частинки розподілялися по деревах або порівну, або таким чином щоб кількість частинок на різних деревах відрізнялися не більш ніж на одну. На відміну від [1], при розрахунку амплітуди необхідно розрахувати суму по всіх можливих розподілах частинок по деревах. Ми поки що цього не робили, але розрахунок такої суми може тільки посилити ті ефекти, про які йде мова далі.

Як видно на рис.15 в широкому інтервалі енергій інтерференційні внески мають суттєвий вклад у парціальний переріз.

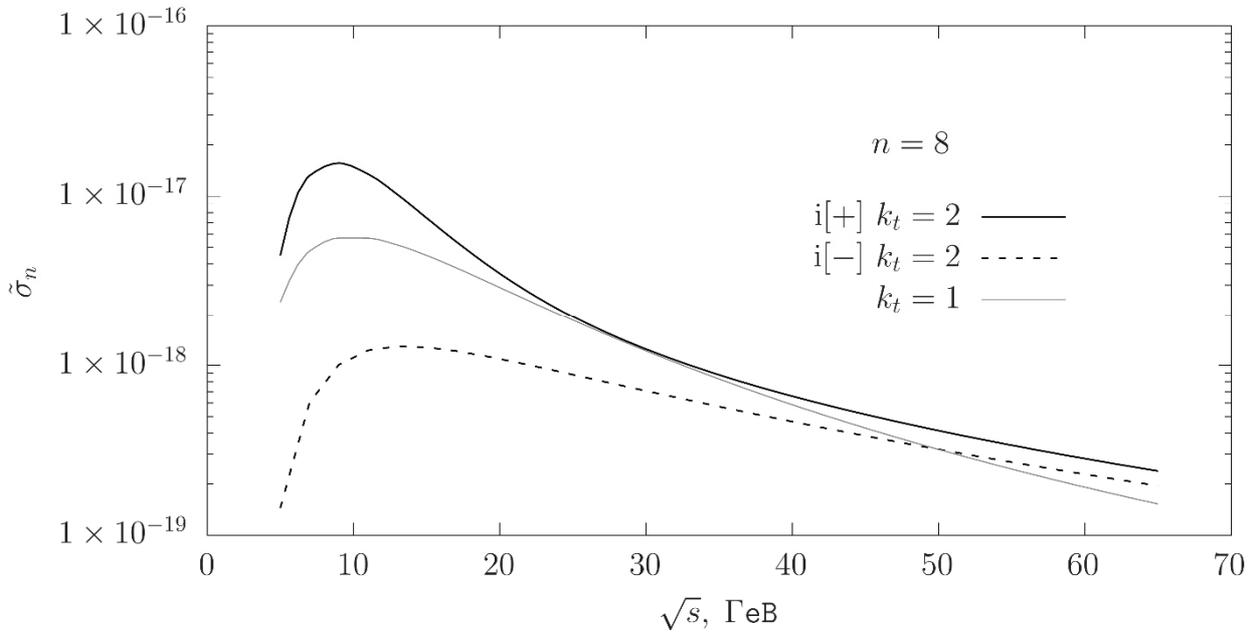

Рис. 15. Залежність внеску у парціальний переріз від діаграм виду рис. 1 з урахуванням (i[+]) та без урахування (i[−]) інтерференційних внесків, що відповідають перестановкам між різними деревами для $n=8$.

Дійсно, для двох дерев, як і для двох груп у попередніх розділах, відстань між точками максимумів двох виразів $A_n(\sqrt{s}, X)$ і $A_n(\sqrt{s}, \hat{\pi}X)$, де $\pi \notin [\varepsilon]$ (позначення відповідають роботі [1]) зростає з ростом енергії. Це призводить до того, що добуток $A_n(\sqrt{s}, X) A_n(\sqrt{s}, \hat{\pi}X)$ у підінтегральному виразі інтерференційного внеску, що відповідає перестановці $\pi$, зменшується на всій області інтегрування. Це добре видно на рис.15 – суцільна та штрихова лінії зближаються з ростом енергії, що свідчить про те, що інтерференційні внески, що відповідають перестановкам між різними деревами, поступово стають менш значними з ростом енергії. При цьому необхідність врахування усіх інтерференційних внесків очевидна – їх врахування призводить до зміни внеску у парціальні перерізи на порядки величини в усьому діапазоні енергій, в якому проводились розрахунки. Такі ж самі висновки можна зробити з результатів, наведених на рис. 16 -17.

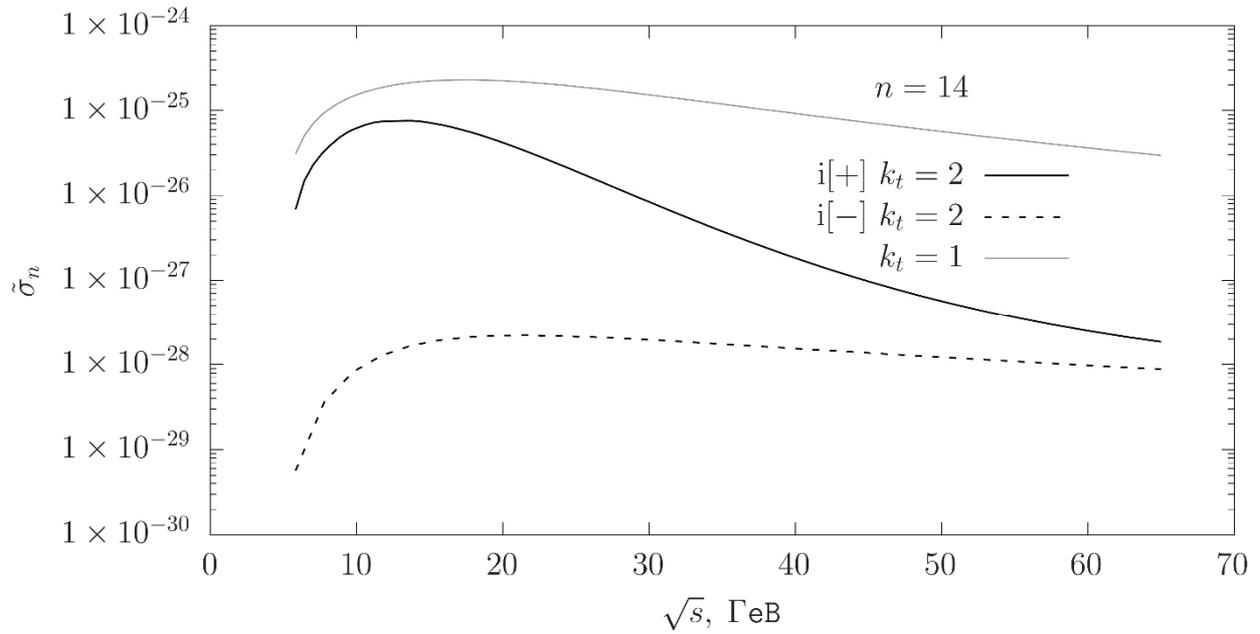

Рис. 16. Залежність внеску у парціальний переріз від діаграм виду рис. 1 з урахуванням (i[+]) та без урахування (i[−]) інтерференційних внесків, що відповідають перестановкам між різними деревами для $n=14$.

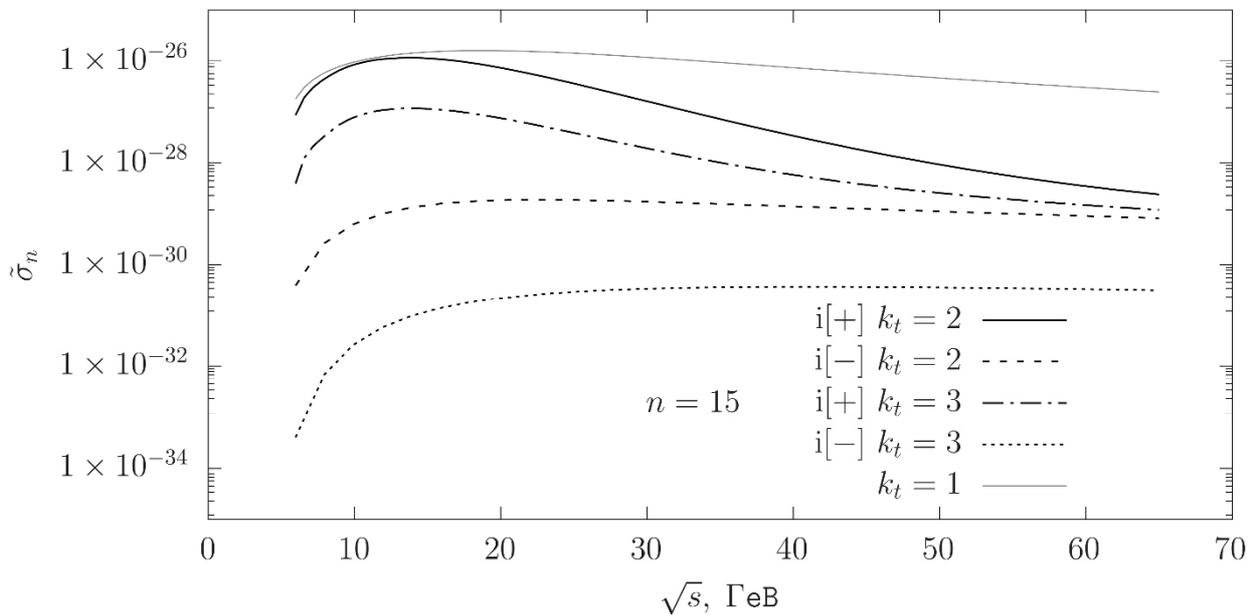

Рис. 17. Залежність внеску у парціальний переріз від діаграм виду рис. 1 з урахуванням (i[+]) та без урахування (i[−]) інтерференційних внесків, що відповідають перестановкам між різними деревами для $n=15$.

Як вже було сказано, це пов'язано з тим, що сума по перестановкам вторинних частинок в середині дерев $A_n(\sqrt{s},X)$ має єдину точку максимуму незалежно від значення енергії. При цьому з ростом енергії цей максимум не стає менш *різким* (формально – не зменшуються власні значення матриці других похідних) внаслідок віддалення точок максимумі доданків суми $A_n(\sqrt{s}, X)$. Враховуючи, що з ростом енергії максимуми двох доданків $A_n(\sqrt{s}, X)$ і $A_n(\sqrt{s}, \hat{\pi} X)$ у сумі для $T_n(\sqrt{s}, X)$ [1] віддаляються все більше, то кожен окремий максимум функції $T_n(\sqrt{s}, X)$ все краще апроксимується відповідною функцією $A_n(\sqrt{s}, \hat{\pi} X)$.

## 7. Можливість застосування наведених результатів для моделювання партонних злив в КХД

В роботі [13] було показано, що діаграми КХД з глюонними петлями, типу наведеної на рис.18а дозволяють розрахунок методом Лапласа. Аналогічний розрахунок був застосований у роботі [14] до діаграм скалярної моделі «фі-три» (рис.18б). В [13] показано, що всі властивості максимумів модулів амплітуд, що відповідають обом діаграмам на рис. 18б є аналогічними. Обидві амплітуди представляються в виді добутку множників, які в одному випадку відповідають петлям, а в другому лініям віртуальних частинок. При цьому множник, що відповідає петлі, в яку втікає певний чотиріімпульс в околі точки максимуму поводиться подібно множнику, що відповідає лінії, по якій переноситься такий самий чотиріімпульс. На рис.18в ці петлі і лінії об'єднані одним прямокутником, що підкреслює їх подібність. Більш конкретно, що означає ця подібність видно з рис. 19. Як показано в [13] множник, який відповідає петлі в околі точки максимуму поводиться як квадратний корінь з множника, що відповідає лінії, по якій переноситься такий самий чотиріімпульс, який втікає в петлю. При цьому в обох випадках $K^2 < 0$ і вираз під коренем є додатним. Таким чином амплітуда, що відповідає петльовій діаграмі на рис 18в в околі точки максимуму поводиться як корінь (монотонна функція) від амплітуди, що відповідає гребінці. Тому ці функції досягатимуть максимуму в одній тій самій точці і властивості максимуму є однаковими. Оскільки в методі Лапласа ми маємо справу з логарифмом модуля амплітуди, то ці логарифми відрізняються лише на постійний множник.

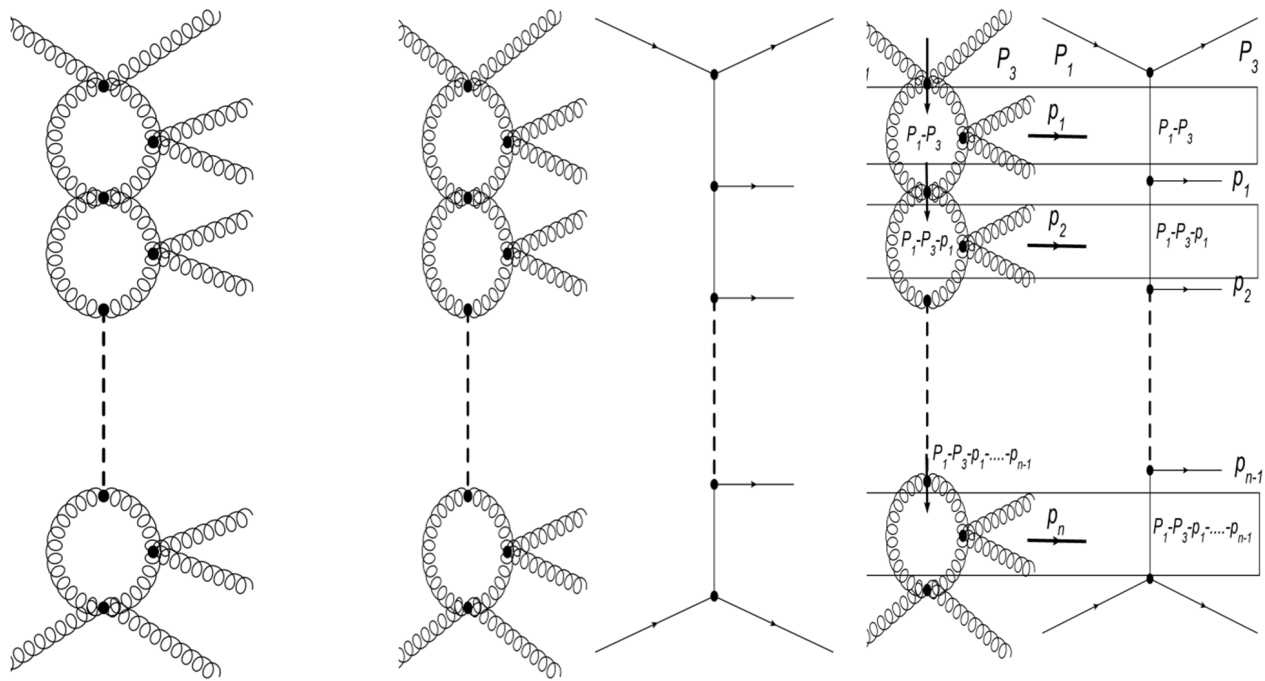

а)                                  б)                                  в)

Рис.18. Діаграма з $n$ глюонними петлями в КХД(а), і подібність між цією діаграмою і гребінкою з $n-1$ вторинною частинкою в скалярній моделі «фі-три» (б). Позначення $p_1, p_2, \ldots, p_n$ означають сумарний чотириімпульс, який уноситься двома глюонами, що приєднуються до відповідної петлі. Кожен множник, що відповідає петлі в околі точки максимуму діаграми КХД подібний до відповідного множника, що відповідає лінії на діаграмі моделі «фі-три» (в).

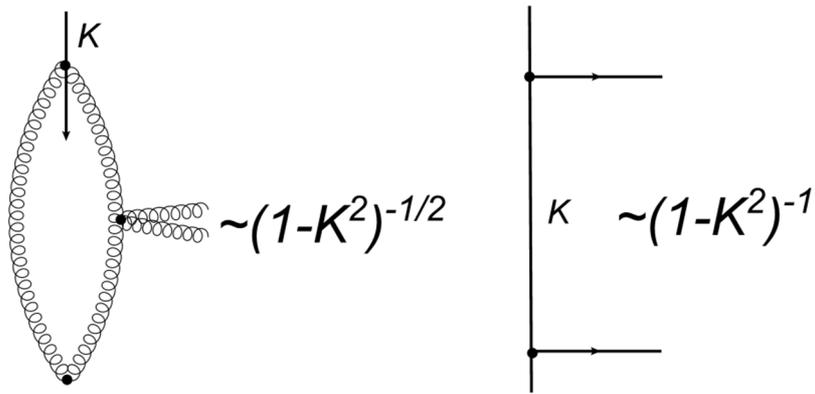

Рис. 19. В околі точки максимуму логарифми множників, що відповідають петлі, в яку втікає чотириімпульс $K$ і лінії гребінки, по якій протікає такий самий чотириімпульс, відрізняються на постійний множник.

Окрім того, в роботі [14] показано, що множник, в амплітуді, що відповідає великій глюонній петлі типу рис,20а досягає максимального значення за рівних чотириімпульсів $p_1, p_2, \ldots, p_n$, які уносяться вихідними глюонами. Тобто такий множник, еквівалентний деревному блоку на рис.1 і описує партонну зливу з факторіальним підсиленням константи зв'язку.

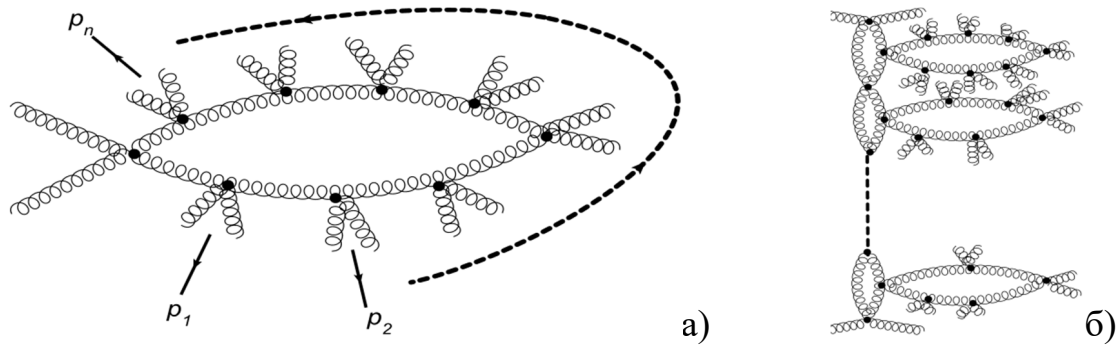

Рис.20 Велика глюонна петля з однаковими чотириімпульсами зовнішніх ліній в точці максимуму (а) і діаграма з такими петлями, еквівалентна діаграмі на рис.1

Комбінуючи ці результати можемо сказати, що діаграми типу рис. 20б описують процес утворення декількох партонних злив. Для таких діаграм, таким чином, можна застосувати метод розрахунку, описаний в роботі [1] і в цій роботі.

## 8. Висновок

При описі процесів з утворенням адронних і партонних злив, зазвичай, не враховується інтерференція між різними зливами. В цій роботі ми показали, що врахування цієї інтерференції є необхідним.

## Література